\documentclass[rnaas]{aastex62}
\usepackage{graphicx,amsmath}

\begin{document}

\title{Spectroscopic Confirmation that 2MASS J07414279$-$0506464 is a Mid-Type L Dwarf}

\correspondingauthor{Michael Cushing}
\email{michael.cushing@utoledo.edu}

\author[0000-0001-7780-3352]{Michael C. Cushing}
\affiliation{Department of Physics and Astronomy, University of Toledo, 2801 W. Bancroft St., Toledo, OH 43606, USA}

\author[0000-0001-6765-6336]{Nicholas Moskovitz}
\affiliation{Lowell Observatory, 1400 West Mars Hill Road, Flagstaff, AZ 86001, USA}

\author[0000-0002-7600-4652]{Annika Gustafsson}
\affiliation{Department of Physics and Astronomy, P.O. Box 6010, Northern Arizona University, Flagstaff, AZ 86011, USA}
\altaffiliation{Lowell Observatory, 1400 West Mars Hill Road, Flagstaff, AZ 86001, USA}







\section{} 

At least one star or brown dwarf has been added to the list of stellar systems that lie within 10 pc of the Sun every year since 2002 \citep{2018arXiv180407377H}.  Motivated by the uneven spatial distribution of the nearest brown dwarfs which suggests that there may be more members of the solar neighborhood \citep{2016A&A...589A..26B},  \citet{2018RNAAS...2a..33S} recently identified three potential nearby ($d\lesssim 20$ pc) L dwarfs using data from various optical and infrared wide-field or all-sky surveys:  2MASS J07555430$-$3259589 with a photometric spectral type of L7.5p that may be a member of the Carina-Near moving group, 2MASS J07414279$-$0506464 (hereafter 2MASS 0741$-$05) with a photometric spectral type of L5 that is resolved in the first Gaia data release as a close ($0\farcs$3) binary, and 2MASS J19251275$+$0700362 (hereafter 2MASS 1925$+$07), with a photometric spectral type of L7.  \citet{2018arXiv180501573F} recently confirmed a spectral type of L7 for 2MASS 1925$+$07 using a moderate-resolution near-infrared spectrum obtained with the TripleSpec instrument on the Palomar 200$''$ telescope \citep{2008SPIE.7014E..30H}.  Here we present a low-resolution, near-infrared spectrum of 2MASS 0741$-$05 and confirm that it is a mid-type L dwarf.

We observed 2MASS 0741$-$05 on 2018 Mar 02 (UT) with the Near-Infrared High Throughput Spectrograph \citep[NIHTS;][]{2011epsc.conf.1823R} mounted on the Lowell Observatory's 4.3 m Discovery Channel Telescope (DCT) located in Happy Jack, AZ.  We used the 1$\farcs$34-wide slit which provides a resolving power of $R\equiv \lambda/\Delta \lambda \approx 150$--35 over the 1--2.4 $\mu$m wavelength range.  We obtained 10, 60 s exposures at two different positions along the 10$''$-long slit.  Flat field exposures and Xe arc exposures were taken for calibration purposes and the A0 Vn star HD 67725\footnote{SIMBAD currently lists the spectral type of HD 67725 as B8 V but we drew our list of potential A0 V stars from SIMBAD in September 2001 when its spectral type was listed as A0 Vn.} was observed for telluric correction and flux calibration purposes.  The data were reduced with the IDL-based Spextool data reduction package \citep{2004PASP..116..362C}.  The package was recently upgraded to reduce data obtained with iSHELL, the new high resolution ($R\approx 70,000$)  infrared spectrograph on the NASA Infrared Telescope Facility \citep{2016SPIE.9908E..84R}.  The NIHTS version of Spextool uses the new iSHELL code to account for the tilt of the slit image relative to the columns of the detector.  Telluric correction and flux calibration were accomplished using the A0 Vn star and the technique described in \citet{2003PASP..115..389V}.  We achieved signal-to-noise ratios of 100, 300, and 400 at the peaks of the $J$, $H$, and $K$ bands, respectively.

We estimate the spectral type of 2MASS 0741$-$05 by comparing our spectrum to spectra of the L dwarf spectral standards of \citet{2010ApJS..190..100K} taken at similar resolving powers ($R\approx$100--300).   Figure 1 shows the NIHTS spectrum of 2MASS 0741$-$05 (black) along with the spectra of the L4, L5, L6, and L7 spectral standards (red).  Spectral types on the \citet{2010ApJS..190..100K} are determined primarily over the 0.9--1.4 $\mu$m wavelength range which suggests 2MASS 0741$-$05 is an L4 or an L5 dwarf.  The longer wavelength portions of the spectrum are best matched to the L5 standard and so we assign a spectral type of L5 to 2MASS 0741$-$05.

\begin{figure}[h]
\begin{center}
\includegraphics[angle=0]{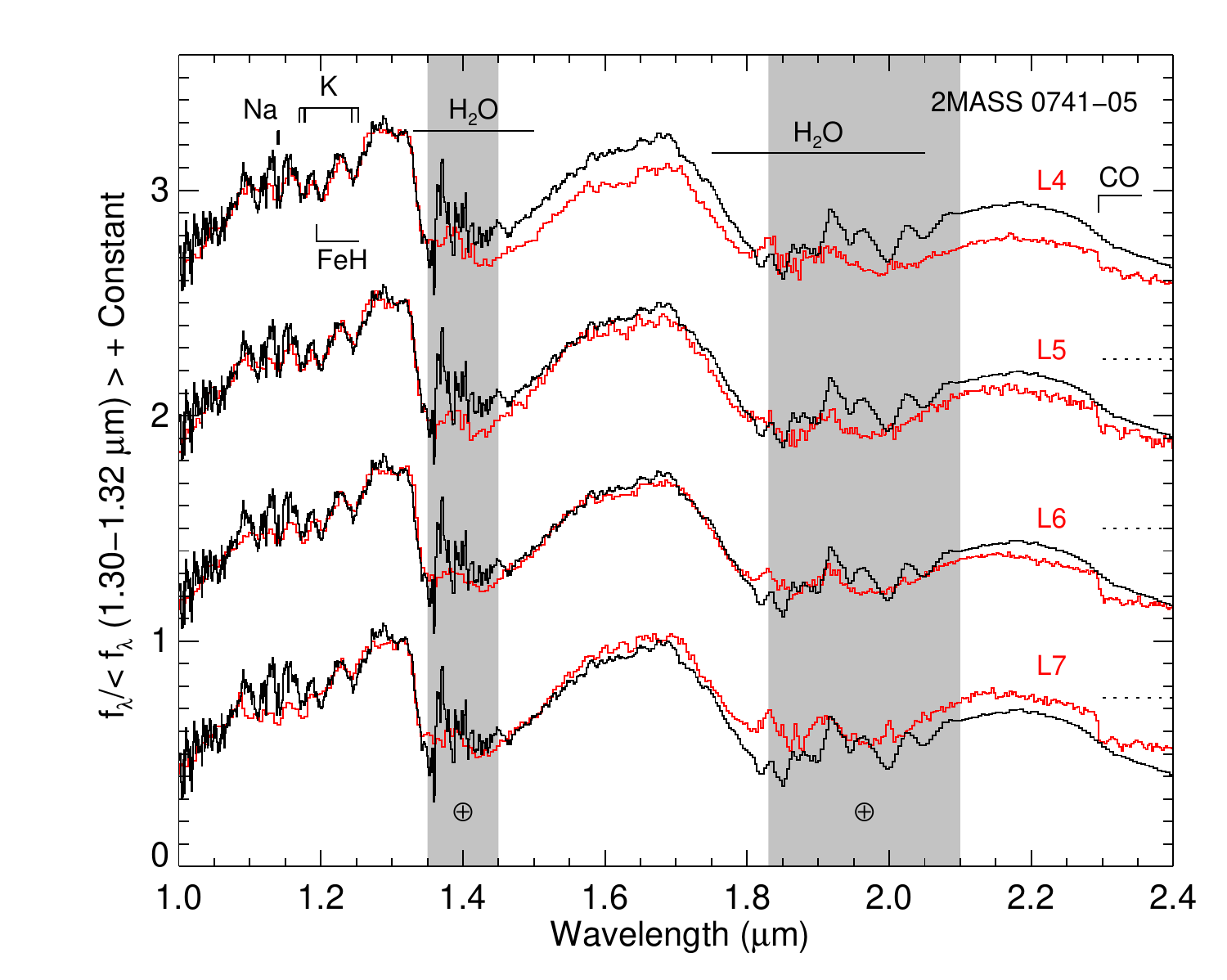}
\caption{1--2.4 $\mu$m spectrum of 2MASS 0471$-$05 (\textsl{black}) along with the spectra of the near-infrared L dwarf spectral standards (\textsl{red}) 2MASS J21580457$–$1550098 (L4), 2MASS J08350622$-$195305 (L5), 2MASS J10101480$-$0406499 (L6), and 2MASS J01033203$-$1935361 (L7) from \citet{2010ApJS..190..100K}.  Prominent atomic and molecular features are indicated as well as regions of strong telluric absorption (grey bars).} 
\end{center}
\end{figure}

\acknowledgments

This research has made use of the SIMBAD database, operated at CDS, Strasbourg, France.  
\facility{Lowell DCT (NIHTS)}, \software{Spextool \citet{2004PASP..116..362C}}

\bibliographystyle{apj}
\bibliography{/Users/mcushing/Science/Papers/ref,/Users/mcushing/Science/Papers/tmp}

\begin{thebibliography}{12}
\expandafter\ifx\csname natexlab\endcsname\relax\def\natexlab#1{#1}\fi

\bibitem[{{Bihain} \& {Scholz}(2016)}]{2016A&A...589A..26B}
{Bihain}, G., \& {Scholz}, R.-D. 2016, \aap, 589, A26

\bibitem[{{Cushing} {et~al.}(2004){Cushing}, {Vacca}, \&
  {Rayner}}]{2004PASP..116..362C}
{Cushing}, M.~C., {Vacca}, W.~D., \& {Rayner}, J.~T. 2004, \pasp, 116, 362

\bibitem[{{Faherty} {et~al.}(2018){Faherty}, {Gagne}, {Burgasser}, {Mamajek},
  {Gonzales}, {Bardalez Gagliuffi}, \& {Marocco}}]{2018arXiv180501573F}
{Faherty}, J.~K., {Gagne}, J., {Burgasser}, A.~J., {Mamajek}, E.~E.,
  {Gonzales}, E.~C., {Bardalez Gagliuffi}, D.~C., \& {Marocco}, F. 2018, ArXiv
  e-prints

\bibitem[{{Henry} {et~al.}(2018){Henry}, {Jao}, {Winters}, {Dieterich},
  {Finch}, {Ianna}, {Riedel}, {Silverstein}, {Subasavage}, \& {Halley
  Vrijmoet}}]{2018arXiv180407377H}
{Henry}, T.~J., {et~al.} 2018, ArXiv e-prints

\bibitem[{{Herter} {et~al.}(2008){Herter}, {Henderson}, {Wilson}, {Matthews},
  {Rahmer}, {Bonati}, {Muirhead}, {Adams}, {Lloyd}, {Skrutskie}, {Moon},
  {Parshley}, {Nelson}, {Martinache}, \& {Gull}}]{2008SPIE.7014E..30H}
{Herter}, T.~L., {et~al.} 2008, in Presented at the Society of Photo-Optical
  Instrumentation Engineers (SPIE) Conference, Vol. 7014, Society of
  Photo-Optical Instrumentation Engineers (SPIE) Conference Series

\bibitem[{{Kirkpatrick} {et~al.}(2010){Kirkpatrick}, {Looper}, {Burgasser},
  {Schurr}, {Cutri}, {Cushing}, {Cruz}, {Sweet}, {Knapp}, {Barman},
  {Bochanski}, {Roellig}, {McLean}, {McGovern}, \&
  {Rice}}]{2010ApJS..190..100K}
{Kirkpatrick}, J.~D., {et~al.} 2010, \apjs, 190, 100

\bibitem[{{Rayner} {et~al.}(2016){Rayner}, {Tokunaga}, {Jaffe}, {Bonnet},
  {Ching}, {Connelley}, {Kokubun}, {Lockhart}, \&
  {Warmbier}}]{2016SPIE.9908E..84R}
{Rayner}, J., {et~al.} 2016, in \procspie, Vol. 9908, Ground-based and Airborne
  Instrumentation for Astronomy VI, 990884

\bibitem[{{Roe} {et~al.}(2011){Roe}, {Dunham}, {Bida}, {Hall}, \&
  {Degroff}}]{2011epsc.conf.1823R}
{Roe}, H.~G., {Dunham}, E.~W., {Bida}, T.~A., {Hall}, J.~C., \& {Degroff}, W.
  2011, in EPSC-DPS Joint Meeting 2011, 1823

\bibitem[{{Scholz} \& {Bell}(2018)}]{2018RNAAS...2a..33S}
{Scholz}, R.-D., \& {Bell}, C.~P.~M. 2018, Research Notes of the American
  Astronomical Society, 2, 33

\bibitem[{{Skrutskie} {et~al.}(2006){Skrutskie}, {Cutri}, {Stiening},
  {Weinberg}, {Schneider}, {Carpenter}, {Beichman}, {Capps}, {Chester},
  {Elias}, {Huchra}, {Liebert}, {Lonsdale}, {Monet}, {Price}, {Seitzer},
  {Jarrett}, {Kirkpatrick}, {Gizis}, {Howard}, {Evans}, {Fowler}, {Fullmer},
  {Hurt}, {Light}, {Kopan}, {Marsh}, {McCallon}, {Tam}, {Van Dyk}, \&
  {Wheelock}}]{2006AJ....131.1163S}
{Skrutskie}, M.~F., {et~al.} 2006, \aj, 131, 1163

\bibitem[{{Vacca} {et~al.}(2003){Vacca}, {Cushing}, \&
  {Rayner}}]{2003PASP..115..389V}
{Vacca}, W.~D., {Cushing}, M.~C., \& {Rayner}, J.~T. 2003, \pasp, 115, 389

\bibitem[{{Wright} {et~al.}(2010){Wright}, {Eisenhardt}, {Mainzer}, {Ressler},
  {Cutri}, {Jarrett}, {Kirkpatrick}, {Padgett}, {McMillan}, {Skrutskie},
  {Stanford}, {Cohen}, {Walker}, {Mather}, {Leisawitz}, {Gautier}, {McLean},
  {Benford}, {Lonsdale}, {Blain}, {Mendez}, {Irace}, {Duval}, {Liu}, {Royer},
  {Heinrichsen}, {Howard}, {Shannon}, {Kendall}, {Walsh}, {Larsen}, {Cardon},
  {Schick}, {Schwalm}, {Abid}, {Fabinsky}, {Naes}, \&
  {Tsai}}]{2010AJ....140.1868W}
{Wright}, E.~L., {et~al.} 2010, \aj, 140, 1868

\end{thebibliography}

\end{document}